\documentclass[a4paper,12pt]{article}
\usepackage{bm}
\usepackage{amsmath}
\usepackage{amssymb}
\usepackage[dvips]{graphicx}
\usepackage[dvips]{psfrag}

\makeatletter
\@addtoreset{equation}{section}
\renewcommand{\theequation}{\thesection.\@arabic\c@equation}
\makeatother
\makeatletter
\renewcommand\appendix{\par
  \setcounter{section}{0}%
  \setcounter{subsection}{0}%
  \gdef\thesection{Appendix \@Alph\c@section }
  \renewcommand{\theequation}
  {\Alph{section}.\arabic{equation}}
}
\makeatother

\def\nn{\nonumber}
\def\lb{\label}
\def\ci{\cite}
\newcommand{\Ref}[1]{(\ref{#1})}
\def\a{\alpha}
\def\b{\beta}
\def\g{\gamma}

\def\th{\theta}
\def\s{\sigma}

\newcommand{\rd}{\mathrm{d}}
\newcommand{\p}{\partial}

\def\bra{\langle}
\def\ket{\rangle}
\def\l{\left}
\def\r{\right}

\def\f{\frac}

\def\bk{\bm k}
\def\bl{\bm l}

\begin{document}

\begin{titlepage}

\vspace*{-15mm}   
\baselineskip 10pt   
\begin{flushright}   
\begin{tabular}{r}    
{\tt KUNS-2518}\\ 
{\tt YITP-14-115}\\
March 26, 2015
\end{tabular}   
\end{flushright}   
\baselineskip 24pt   
\vglue 10mm   

\begin{center}
{\Large\bf Phenomenological Description of \\
the Interior of the Schwarzschild Black Hole }

\vspace{8mm}   

\baselineskip 18pt   

\renewcommand{\thefootnote}{\fnsymbol{footnote}}

Hikaru~Kawai$^a$\footnote[2]{hkawai@gauge.scphys.kyoto-u.ac.jp} and 
Yuki~Yokokura$^b$\footnote[4]{yuki.yokokura@yukawa.kyoto-u.ac.jp}

\renewcommand{\thefootnote}{\arabic{footnote}}

\vspace{5mm}   

{\it  
 $^a$ Department of Physics, Kyoto University, 
 Kitashirakawa, Kyoto 606-8502, Japan \\
 $^b$ Yukawa Institute for Theoretical Physics, Kyoto University, Kyoto 606-8502, Japan
}
  
\vspace{10mm}   

\end{center}

\begin{abstract}
We discuss a sufficiently large 4-dimensional Schwarzschild black hole which is in equilibrium with a heat bath. 
In other words, we consider a black hole which has grown up from a small one in the heat bath adiabatically. 
We express the metric of the interior of the black hole in terms of two functions: 
One is the intensity of the Hawking radiation, 
and the other is the ratio between the radiation energy and the pressure in the radial direction. 
Especially in the case of conformal matters we check that it is a self-consistent solution 
of the semi-classical Einstein equation, $G_{\mu\nu}=8\pi G \bra T_{\mu\nu}\ket$. 
It is shown that the strength of the Hawking radiation is proportional to the c-coefficient, 
that is, the coefficient of the square of the Weyl tensor in the 4-dimensional Weyl anomaly. 
\end{abstract}

\baselineskip 18pt   

\end{titlepage}

\newpage
\section{Motivation}
A large black hole evaporates slowly in the vacuum \ci{Hawking}, and 
it also becomes equilibrium with the heat bath of the Hawking temperature \ci{G-H}.
Especially we can grow a small black hole to a large one adiabatically, and vice versa, 
by changing the temperature of the heat bath properly\footnote{In such process we also need to change the size of the heat bath properly 
because a black hole has the negative heat capacity.}. 
Then, the inside of the black hole is expected to be stuffed with matters and radiations, 
and it should make sense to consider the metric inside the black hole. 
In fact, we show that the whole history of the black hole including formation and evaporation can be described by the Schwarzschild coordinate globally. 
Furthermore, we also demonstrate that the energy is distributed in the whole region inside the black hole, 
while the pressure is totally anisotropic: $\bra T^\th{}_\th \ket \gg \bra T^r{}_r \ket$. 
In this sense our picture is different from membrane models \ci{membrane} and fluid models. 

In particular, we can consider the inside metric of a Schwarzschild black hole with radius $a$ which is in equilibrium with a heat bath. 
Because it is static and spherically symmetric, we can parametrize it as
\begin{equation}\lb{ansatz}
\rd s^2 = - \f{1}{B(r)}e^{A(r)}\rd t^2+ B(r) \rd r^2 + r^2 \rd \Omega^2,~~~{\rm for}~~r<a~.
\end{equation}
In general, $A(r)$ and $B(r)$ depend on the radius $a$. 
However, we can show that such metric is nearly independent of the radius $a$, as follows. 
First, we discuss the relaxation time of the black hole. 
As is well-known, the time scale of the evaporation of a black hole with radius $a$ is of order $\f{a^3}{l_p^2}$. 
Here $l_p$ is the Planck length, $l_p^2\equiv G\hbar$. 
Therefore, it is natural to postulate that 
the relaxation time of the black hole is at most of order $\f{a^3}{l_p^2}$. 
In other words, a change slower than this time scale can be regarded as an adiabatic change. 
On the other hand, as we will see in the next section, 
the redshift factor inside the black hole is exponentially large so that 
the time inside is almost frozen. 
Therefore, if we consider a process in which a black hole with $a$ is adiabatically shrunk to $a'$ in the heat bath, 
the metric at $r<a'$ changes very little. 
It is because the inside time corresponding to the outside time scale of order $\f{a^3}{l_p^2}$ is very small due to the exponentially large redshift. 
This means that the metric at $r<a'$ of the black hole with radius $a'$ is the same as that of the black hole with radius $a$. 
Thus, as a first trial, we assume that 
the metric $A(r)$ and $B(r)$ do not depend on the size of the black hole. 

Using the same argument we can show that 
the inside of the black hole in the vacuum which has been made adiabatically has the same metric as that in equilibrium. 
In fact, if we put an evaporating black hole into a heat bath, 
it will become equilibrium in the relaxation time, which we have assumed at most of order $\f{a^3}{l_p^2}$. 
Again the metric inside is almost frozen, which indicates that two metrics are the same. 

In the following part we will give the metric \Ref{ansatz} in terms of two phenomenological functions, 
and see that the above discussion is consistent. 

\section{Determination of $A(r)$ and $B(r)$}
In the vacuum a black hole with a large mass $m=\f{a}{2G}$ compared with the Planck mass evaporates slowly according to \ci{Hawking}
\begin{equation}\lb{da}
\f{\rd a}{\rd t}=-\f{2\s(a)}{a^2}~.
\end{equation}
Here $\s(a)/l_p^2$ is a quantity of order one. 
$\s(a)$ describes the intensity of the Hawking radiation, and 
depends on the detail of the theory, such as the number of fields. 
In general $\s(a)$ varies with $a$ slowly compared with $l_p$; $\f{\rd \s}{\rd a}l_p \ll \s$. 

In order to determine $A(r)$ and $B(r)$, we first investigate the surface of the Schwarzschild black hole with radius $a$. 
Here the surface means 
the boundary between the empty space and the region filled with the matters and radiations. 
We denote the radius of the surface by $R(a)$. 
In other words, the region $r<R(a)$ is filled with matters and radiations, while the region $r>R(a)$ is empty. 
From the following argument we can show that 
the notion of the surface is universal and plays a crucial role in the quantum mechanical description of black holes. 

To do it, let us consider the motion $r(t)$ of a test particle approaching to the evaporating black hole. 
We assume that outside the black hole the metric is given by 
\begin{equation}\lb{Sch}
\rd s^2 = - \f{r-a(t)}{r}\rd t^2 + \f{r}{r-a(t)}\rd r^2 + r^2 \rd \Omega^2~. 
\end{equation}
If $r(t)$ is sufficiently close to $a(t)$, 
it is determined by 
\begin{equation}\lb{r_t}
\f{\rd r(t)}{\rd t}=-\f{r(t)-a(t)}{r(t)}~,
\end{equation}
no matter what mass or angular momentum the particle has. 
From this we see that the particle approaches the radius $a$ in the time scale of $a$. 
However, in this case the radius $a(t)$ itself is slowly shrinking by the Hawking radiation. 
Therefore, the particle cannot catch up with the radius $a$ completely. 
Instead, $r(t)$ is always apart from $a(t)$ by $-a\f{\rd a}{\rd t}$. 
Actually, the solution of \Ref{r_t} is approximately given by 
\begin{align}\lb{R0}
r(t)&\approx a-a\f{\rd a}{\rd t}+Ce^{-\f{t}{a}} \nn \\
 &= a+\f{2\s}{a}+Ce^{-\f{t}{a}}
\end{align}
where $C$ is a positive constant and we have used \Ref{da} to obtain the second line\footnote{Putting $r(t)=a(t)+\Delta r(t)$ in \Ref{r_t} and assuming $\Delta r(t)\ll a(t)$, 
we have $\f{\rd \Delta r(t)}{\rd t} = -\f{\Delta r(t)}{a(t)}-\f{\rd a(t)}{\rd t}$. 
The general solution is given by 
$\Delta r(t)=C_0 e^{-\int^t_{t_0}\rd t' \f{1}{a(t')}} + \int^t_{t_0}\rd t' \l(-\f{\rd a}{\rd t}(t') \r)e^{-\int^{t}_{t'}\rd t''\f{1}{a(t'')}}$, 
where $C_0$ is an integration constant. 
Here, because of \Ref{da}, $a(t)$ and $\f{\rd a}{\rd t}(t)$ change so slowly that they can be considered constant in the time scale ${\cal O}(a)$. 
Then, the second term can be evaluated as 
$\int^t_{t_0}\rd t' \l(-\f{\rd a}{\rd t}(t') \r)e^{-\int^{t}_{t'}\rd t''\f{1}{a(t'')}}
\approx -\f{\rd a}{\rd t}(t) \int^t_{t_0}\rd t' e^{-\f{t-t'}{a(t)}}=-\f{\rd a}{\rd t}(t) a(t)(1-e^{-\f{t-t_0}{a(t)}})$. 
Therefore, we reach $\Delta r(t)\approx C_0 e^{-\f{t-t_0}{a(t)}} -\f{\rd a}{\rd t}(t)a(t) (1-e^{-\f{t-t_0}{a(t)}})$, 
which leads to \Ref{R0}.}. 
This result indicates that 
any particle approaches 
\begin{equation}\lb{R}
R(a)\equiv a+\f{2\s}{a}
\end{equation}
within the time scale of ${\cal O}(a)$. 
Therefore, 
any black hole with the Schwarzschild radius $a$ has such a universal surface at $r=R(a)$, 
no matter how it has been formed. 
Thus, it is natural to postulate that $A(r)$ and $B(r)$ in \Ref{ansatz} can be used up to $r=R(a)$: 
\begin{equation}\lb{ansatz2}
\rd s^2 = - \f{1}{B(r)}e^{A(r)}\rd t^2+ B(r) \rd r^2 + r^2 \rd \Omega^2,~~~{\rm for}~~r\leq R(a)~.
\end{equation}
Note that $R(a)$ is slightly larger than $a$ due to the quantum effect, and the horizon no longer exists. 

Now we can determine $B(r)$ as follows. 
First $g_{rr}$ on the surface is obtained from \Ref{Sch} by setting $r=R(a)$: 
\begin{equation}
g_{rr}|_{r=R(a)}=\f{R(a)}{R(a)-a}=\f{R(a)a}{2\s(a)}\approx \f{R(a)^2}{2\s(R(a))}~. 
\end{equation}
In the last expression we have replaced $a$ with $R(a)$ 
because $\f{2\s}{a}$ is much less than $a$ for a large black hole, $a\gg l_p$. 
This can be directly compared with $B(R(a))$ in \Ref{ansatz2} because the radial coordinate $r$ 
is uniquely fixed in the Schwarzschild coordinate \Ref{Sch}: 
$B(R(a))=\f{R(a)^2}{2\s(R(a))}~.$
Because this result holds for any $a$, and we have postulated that $A(r)$ and $B(r)$ do not depend on $a$,  
we find that the function $B(r)$ is universally given by the function $\s(r)$ as 
\begin{equation}\lb{B}
B(r)=\f{r^2}{2\s(r)}~.
\end{equation}

Next, in order to fix $A(r)$, 
we consider the energy-momentum flow inside the Schwarzschild black hole which is in equilibrium with a heat bath. 
It is characterized by the following time-reversal symmetric equations: 
\begin{equation}\lb{T1}
-\bra T^{\mu \nu} \ket k_\nu=\eta (l^\mu+f(r) k^\mu),~~~-\bra T^{\mu \nu}\ket l_\nu=\eta (k^\mu+f(r) l^\mu)~,
\end{equation}
where $f(r)$ is a quantity of order one and varies slowly compared with $l_p$, 
and $\bl$ and $\bk$ are the radial outgoing and ingoing null vectors, respectively:
\begin{equation}\lb{l-k}
\bl = e^{-\f{A}{2}}\p_t+\f{1}{B}\p_r ,~~~\bk = e^{-\f{A}{2}}\p_t-\f{1}{B}\p_r ~.
\end{equation}
They transform under time reversal as $(\bl,\bk)\rightarrow (-\bk,-\bl)$. 
These equations can be rewritten as   
\begin{equation}\lb{T2}
\bra T^{\bk \bk}\ket:\bra T^{\bl \bk}\ket=1:f,~~~\bra T^{\bk \bk}\ket=\bra T^{\bl \bl}\ket~,
\end{equation}
where $T^{\bk \bk}$ stands for $T^{\mu \nu}k_\mu k_\nu$, and so on. 

Now we discuss the physical meaning of $f$. 
The vector $P^\mu=\bra T^{\mu \bk}\ket$ at $r$ represents the energy-momentum flow through the ingoing lightlike spherical surface $S$ of radius $r$. 
We can regard $S$ as the surface of an evaporating black hole 
because the radius of $S$ satisfies \Ref{r_t}. 
Then, $P^\mu$ describes the Hawking radiation from the black hole. 
If the radiated particle is massless and propagates outward along the radial direction without scattering, 
$P^\mu$ should be parallel to $l^\mu$, which means $f=0$. 
Therefore, $f$ represents the deviation from this ideal situation. 
If the radiated particle is massive, $P^\mu$ is timelike, and we have $f>0$. 
Even when the particle is massless, $f$ can become non-zero if the particle is scattered in the ingoing direction. 
This is because such a scattered particle comes back to the surface in the time scale of $a$, as in \Ref{R0}, 
and produces an energy-momentum flow along $\bk$ direction. 
Finally, we point out that \Ref{T2} can be expressed in terms of  
the ratio between the energy density $-\bra T^t{}_t \ket$ and the pressure in the radial direction $\bra T^r{}_r \ket$: 
\begin{equation}\lb{T3}
\f{\bra T^r{}_r \ket}{- \bra T^t{}_t \ket}=\f{1-f}{1+f}~.
\end{equation}

Once $f(r)$ is given, we can determine $A(r)$. 
Using \Ref{T3} and the Einstein equation, we have
\begin{equation}\lb{dA}
\f{2}{1+f} =\f{G^r{}_r}{-G^t{}_t}+1=\f{r\p_r A}{B-1+r\p_r \log B}\approx \f{r\p_r A}{B}~. 
\end{equation}
In the last equation, we have used $B\gg 1$ and $B\gg r\p_r \log B$ for $r \gg l_p$, which can be easily shown from \Ref{B}.  
Thus, we obtain 
\begin{equation}\lb{A}
A(r)=\int^r_{r_0}\rd r'\f{r'}{(1+f(r'))\s(r')}~,
\end{equation}
where $r_0$ is a reference point. 

Now, by connecting \Ref{ansatz2} and the Schwarzschild metric at the surface, 
we can write down the metric of the black hole with radius $a_0$ which is in equilibrium with the heat bath: 
\begin{equation}\lb{sta_metric}
\rd s^2=\begin{cases}
-\f{2\s(r)}{r^2} e^{- \int^{R_0}_r \rd r' \f{r'}{(1+f(r'))\s(r')}} \rd t^2 + \f{r^2}{2\s(r)} \rd r^2 + r^2 \rd \Omega^2,~~{\rm for}~~r\leq R_0~,\\
 - \f{r-a_0}{r}\rd t^2 + \f{r}{r-a_0}\rd r^2 + r^2 \rd \Omega^2,~~{\rm for}~~r\geq R_0~,
\end{cases}
\end{equation}
where $R_0=a_0 + \f{2\s(a_0)}{a_0}$. 
This metric is continuous at $r=R_0$ \footnote{The inside metric of \Ref{sta_metric} does not exist in the classical limit 
$\hbar \rightarrow 0$ because $\s(r)=\hbar G {\cal O}(1)$.}. 

Similarly, we can construct the metric of the evaporating black hole in the vacuum. 
In order to do that, 
we first rewrite the metric \Ref{ansatz2} in the Eddington-Finkelstein-like coordinates as
\begin{equation}\lb{EF}
\rd s^2 = - e^{\f{A(r)}{2}}\l(\f{1}{B(r)}e^{\f{A(r)}{2}}\rd u+2\rd r \r)\rd u + r^2 \rd \Omega^2~,
\end{equation}
where we have introduced $u$-coordinate by $\rd u = \rd t - B(r)e^{-\f{A(r)}{2}}\rd r$. 
Then, we obtain the metric by connecting \Ref{EF} to the Vaidya metric \ci{Vaidya} along the null surface $S$: 
\begin{equation}\lb{eva_metric}
\rd s^2=\begin{cases}
-e^{- \int^{R_0(u)}_r \rd r' \f{r'}{2(1+f(r'))\s(r')}} \l(\f{2\s(r)}{r^2} e^{- \int^{R_0(u)}_r \rd r' \f{r'}{2(1+f(r'))\s(r')}} \rd u 
+ 2\rd r \r) \rd u + r^2 \rd \Omega^2,~~{\rm for}~~r\leq R_0(u)~,\\
 - \f{r-a_0(u)}{r}\rd u^2 -2\rd r \rd u + r^2 \rd \Omega^2,~~{\rm for}~~r\geq R_0(u)~,
\end{cases}
\end{equation}
where $R_0(u)=a_0(u) + \f{2\s (a_0(u))}{a_0(u)}$. 
This metric is continuous at $r=R_0(u)$. 
The spacetime described by this metric has the same topological structure as the flat Minkowski space \ci{KMY}.

\section{Consistency checks}
We check the consistency from various points of view. 
First let us check that the redshift factor inside the black hole is indeed exponentially large.  
The $tt$-component of \Ref{sta_metric} behaves as  
$-g_{tt}\sim \exp \l(-\f{a_0}{(1+f)\s}(R_0-r)-2\log \f{a_0}{\sqrt{2\s}} \r)$ slightly below the surface, $r \lesssim R_0$. 
Here we have used the fact that $\s(r)/l_p^2$ and $f(r)$ are quantities of order one, and thus small compared to $r/l_p$. 
This means that the time flows only in the near-surface region with the width of ${\cal O}\l( \f{l_p^2}{a_0}\r)$, 
and the time is frozen in the deeper region, as we have assumed.  

Next we examine the validity of the use of the Einstein equation. 
From \Ref{sta_metric} we can estimate the geometrical invariants for $l_p \lesssim r \leq R_0$ as 
\begin{equation}\lb{curve}
R,~\sqrt{R_{\mu\nu}R^{\mu\nu}},~\sqrt{R_{\mu\nu\a\b}R^{\mu\nu\a\b}}\sim \f{1}{(1+f)^2 \s}~.
\end{equation}
This  means that if the condition 
\begin{equation}\lb{large_N}
\s(1+f)^2\gg l_p^2~
\end{equation}
is satisfied, the curvature is small compared to $l_p^{-2}$, and 
we can use the Einstein equation without introducing the higher-derivative corrections in the entire spacetime. 
In general, it can be realized if we have sufficiently many fields and renormalize the coefficients of the higher-curvature terms to ${\cal O}(1)$ (not proportional to the number of fields). 
Although this field-theoretic approach does not apply to the small region $0\leq  r \lesssim l_p$, 
the curvature at $r\approx l_p$ is smaller than the Planck scale as \Ref{curve}, and 
dynamics in such a small region would be resolved by string theory. 
In this sense, this metric does not have a singularity. 

Then, we investigate the behavior of the energy-momentum tensor inside the black hole. 
They can be evaluated from \Ref{sta_metric} for $r\gg l_p$ as 
\begin{equation}\lb{p}
-\bra T^t{}_t \ket=\f{1}{8\pi G}\f{1}{r^2},~~~
\bra T^\th{}_\th \ket =\f{1}{8\pi G} \f{1}{2(1+f)^2\s}~.
\end{equation}
The energy density $-\bra T^t{}_t \ket$ is positive definite, and it gives the mass of the black hole correctly:
\begin{equation}
m=4\pi \int^R_0 \rd r' r'^2 (-\bra T^t{}_t \ket)\approx \f{a}{2G}~.
\end{equation}
However, the dominant energy condition is broken. 
This is because the pressure in the angular direction $\bra T^\th{}_\th \ket$ is much larger than $-\bra T^t{}_t \ket$ and $\bra T^r{}_r \ket$. 
The existence of large $\bra T^\th{}_\th \ket$ makes a significant difference from the 2-dimensional models \ci{RST}. 
Such a large pressure seems mysterious, but as we will see in the next section, 
it can be naturally understood by the 4-dimensional Weyl anomaly. 
This large pressure leads to drastic anisotropy, and the inside is not a usual fluid, but should be 
viewed as a multi-layer structure\footnote{Note that a similar picture is discussed in the context of black star \ci{black star}.}. 
This picture is consistent with the idea of firewall \ci{firewall} in the sense that 
there is the intense energy density near the surface of the black hole.

Finally, we check that the energy-momentum flow $P^\mu=\bra T^{\mu \bk}\ket$ through the ingoing lightlike spherical surface $S$ 
is consistent with the strength of the Hawking radiation $\s(r)$. 
The total energy flux measured by the local time is given by
\begin{equation}\lb{J_P}
J \equiv 4\pi r^2 P^\mu u_\mu=4\pi r^2 \f{1}{B}(-\bra T^t{}_t \ket)=\f{\s}{Gr^2}~,
\end{equation}
where $\bm u =e^{-\f{A}{2}}\p_t$, and at the last equation we have used \Ref{B} and \Ref{p}. 
This is consistent with \Ref{da}, from which we have $J=-\f{\rd m}{\rd t}=\f{\s}{Ga^2}$.

\section{The case of conformal matters}
As an example of the metric \Ref{sta_metric}, we discuss the case of conformal matters, 
and show explicitly that the metric \Ref{sta_metric} is  a self-consistent solution of the semi-classical Einstein equation, $G_{\mu\nu}=8\pi G \bra T_{\mu\nu} \ket$. 
From the the Weyl anomaly \ci{Duff1}, we have
\begin{equation}\lb{anomaly}
G^\mu{}_\mu=8\pi G \bra T^\mu{}_\mu \ket=\g {\cal F}- \a {\cal G} +\f{2}{3} \b \Box  R~, 
\end{equation}
where ${\cal F}\equiv C_{\mu\nu\a\b}C^{\mu\nu\a\b}$ and ${\cal G}\equiv R_{\mu\nu\a\b}R^{\mu\nu\a\b}-4R_{\mu\nu}R^{\mu\nu}+R^2$. 
We have introduced the reduced notations
\begin{equation}\lb{gamma}
\g\equiv 8\pi G \hbar c,~~~\a\equiv 8\pi G \hbar a,~~~\b\equiv 8\pi G \hbar b~,
\end{equation}
where $c,~a,~b$ are the coefficients in the Weyl anomaly. 
This equation together with the assumption \Ref{T2} determines $A(r)$ and $B(r)$ as follows. 

Here we assume that $A(r)$ and $B(r)$ are large quantities of the same order as in \Ref{B} and \Ref{A}:
\begin{equation}\lb{AB}
A(r)\sim B(r)\gg 1~.
\end{equation}
In order to examine what terms dominate in \Ref{anomaly} for $r\gg l_p$, 
we replace $A$, $B$ and $r$ with $\mu A$, $\mu B$ and $\sqrt{\mu} r$, respectively, 
and pick up the terms with the highest powers of $\mu$. 
Then we evaluate \Ref{anomaly} as \footnote{As we mentioned below \Ref{large_N}, we assume that the coefficients of the higher-curvature terms are 
renormalized to ${\cal O}(1)$. 
However, $\a$ and $\g$ can be large because they are not changed by counter terms. }
\begin{equation}\lb{mu_eq}
\f{A'^2}{2B}+\cdots=\g\l(\f{A'^4}{12 B^2}+\cdots  \r)-\a \l(-\mu^{-1}\f{2A'^2}{r^2B}+\cdots \r)+\f{2}{3}\b\l(\mu^{-1}\l[\f{A'^3B'}{4B^3}-\f{A'^2A''}{2B^2} \r]+\cdots\r)~.
\end{equation}
Therefore, under the assumption $\mu \gg 1$, \Ref{mu_eq} becomes $\f{A'^2}{2B}=\g\f{A'^4}{12 B^2}$, that is, 
\begin{equation}\lb{B2}
B=\f{\g}{6}A'^2 ~.
\end{equation}
By combining this equation with \Ref{dA}, which is the consequence of \Ref{T2}, 
$\f{2}{1+f}=\f{r\p_r A}{B}~$,
we obtain 
\begin{equation}\lb{AB2}
A'=\f{3(1+f)r}{\g},~~~B=\f{3(1+f)^2r^2}{2\g}~.
\end{equation}
We expect that the dimensionless function $f(r)$ is a constant in the case of conformal fields. 
Then, we can obtain easily the inside metric of the Schwarzschild black hole. 
Note that this solution depends only on two parameters: 
One is the $c$-coefficient, which is determined by the matter content, 
and the other is the constant $f$, which depends on the detail of the dynamics.

By comparing the second equation in \Ref{AB2} and \Ref{B}, we find 
\begin{equation}\lb{s}
\s=\f{\g}{3(1+f)^2}~.
\end{equation}
Now the condition for the curvature to be small \Ref{large_N} becomes $\g \gg l_p^2$, that is, 
\begin{equation}\lb{c}
8\pi  c\gg1~.
\end{equation}
It is interesting that the strength of the Hawking radiation is proportional 
to the $c$-coefficient of the 4-dimensional Weyl anomaly\ci{anomaly} \ci{anomaly2}. 
The positive Hawking radiation ensures the positivity of the $c$-coefficient \footnote{This is consistent with a discussion based on spectral representation \ci{spectral}.}. 

Finally, we mention the origin of the strong pressure in the angular direction. 
From \Ref{p} we can show $\bra T^\th{}_\th \ket=\f{1}{2}\bra T^\mu{}_\mu \ket$ for $r\gg l_p$. 
This means that the large pressure naturally comes from the Weyl anomaly. 

\section{Energy flow inside the black hole}
In this section, we discuss how the energy flow is created or annihilated inside the black hole. 
First we consider the outgoing energy flow \Ref{J_P}. 
The energy flow created in the thin spherical region between $r-\Delta r$ and $r$ is given by 
$\Delta J(r)=J(r)-J(r-\Delta r)\f{e^{A(r-\Delta r)}}{e^{A(r)}}\approx e^{-A(r)}\p_r [e^{A(r)}J(r)] \Delta r~$.
Here the factor $e^{A}$ represents the square of the redshift factor $e^{\f{A}{2}}$ 
because $J$ is energy per unit time. 
This equation together with \Ref{A} and \Ref{J_P} gives 
$\f{\Delta J}{\Delta r}=\f{1}{G(1+f)r}$
for $r\gg l_p$, 
which means that 
the outgoing energy is produced at each region and increases as it goes outward. 

Similarly, we can consider the ingoing energy flow. 
From the time reversal we see that it decreases as it goes inward. 
This can be interpreted that the ingoing energy of the matter fields is reduced by 
the negative energy generated inside the black hole \ci{BD}. 
However, the positive energy brought by the ingoing matter is greater than the negative energy 
so that the total energy density is positive everywhere as we have seen in \Ref{p}. 

Therefore, in order to understand the information problem \ci{Hawking2,Bekenstein}, 
it should be important to consider interaction among the ingoing positive energy of the infalling matter, 
the ingoing negative energy, and the outgoing positive energy created inside the black hole. 

\section{Conclusions and Discussions}
If we start with the fact that black holes evaporate, 
we can obtain a self-consistent picture of the spacetime. 
We have considered Schwarzschild black holes which are grown up adiabatically in the heat bath, 
and constructed the interior metric \Ref{sta_metric}. 
It is expressed in terms of two physical quantities: One is the intensity of the Hawking radiation $\s(a)$, and the other is the ratio 
between the energy density and the pressure in the radial direction $f(r)$. 
This is the self-consistent solution of $G_{\mu\nu}=8\pi G \bra T_{\mu\nu} \ket$, which has neither horizon nor singularity. 
The Planck scale corrections can be ignored 
as long as we have sufficiently many fields and renormalize the coefficients of the higher-curvature terms to ${\cal O}(1)$. 
There is large pressure in the angular direction inside the black hole, 
which is naturally understood by the Weyl anomaly in the case of conformal matters. 
Because of that, the dominant energy condition is violated. 

We have also constructed the metric of an evaporating black hole in the vacuum by connecting the inside metric to the Vaidya metric on the ingoing lightlike spherical surface. 
The resulting metric \Ref{eva_metric} has neither a singularity nor a trapped region. 
In our analysis we have considered black holes which are formed adiabatically. 
On the other hand, it is known that an apparent horizon would appear in the generic process of the black hole formation \ci{works}. 
Even in such cases we expect that the dominant energy condition is broken by a mechanism similar to ours, 
and the singularity theorem is avoided \ci{H-E}. 
Because the time scale of the black hole evaporation is the same as the postulated relaxation time, 
the general black hole is expected to approach to a structure that is not very far from \Ref{sta_metric}.

Although we have not discussed the Hawking temperature in this letter, 
we can use the same analysis as in \ci{KMY}. 
We can show that the number of the particles at $r$ takes the form of the Planck-like distribution 
with the temperature $T(r)=\f{\hbar}{4\pi r}$. 

In this paper we have proposed a phenomenological description of the black hole, 
which can be consistently realized by field-theoretic models. 
It gives a simple picture that black holes are no different from the ordinary objects. 
In particular, black holes can be created and annihilated 
without affecting the global structure of the Minkowski space. 
One of the biggest mysteries of the black hole is where the entropy is stored, 
which is interesting also from the viewpoint of quantum information. 
Since the entire metric is determined in our picture, 
we should be able to solve it by investigating 
how the energy and information of matters are transferred into radiation through a black hole. 
We also have seen that our picture becomes consistent if the theory contains many fields. 
It would be interesting to study this effect in the context of string theory, 
which could give some new insight into the Planck scale physics.

%
\section{Acknowledgments}
The authors thank M. Hotta for valuable discussion. 
Y. Y. is partially supported by a Grant-in-Aid of the MEXT Japan for Scientific Research (No. 25287046).
He was also supported by the JSPS Research Fellowship for Young Scientists. 



\begin{thebibliography}{99}
\bibitem{Hawking} 
  S.~W.~Hawking,
  Commun.\ Math.\ Phys.\  {\bf 43}, 199 (1975)  [Erratum-ibid.\  {\bf 46}, 206 (1976)].  

\bibitem{G-H} 
  G.~W.~Gibbons and S.~W.~Hawking,
  Phys.\ Rev.\ D {\bf 15}, 2752 (1977).  


\bibitem{membrane}
K.~S.~Thorne, (Ed.), R.~H.~Price, (Ed.) and D.~A.~Macdonald, (Ed.), \textit{``Black Holes: The Membrane Paradigm"}, NEW HAVEN, USA: YALE UNIV. PR. (1986); 
M.~Saravani, N.~Afshordi and R.~B.~Mann, arXiv:1212.4176.




\bibitem{Vaidya}
P.~C.~Vaidya,\ Proc.\ Indian\ Acad.\ Sci.\ A {\bf 33}, 264 (1951). 

\bibitem{KMY} 
  H.~Kawai, Y.~Matsuo and Y.~Yokokura,
  Int.\ J.\ Mod.\ Phys.\ A {\bf 28}, 1350050 (2013)
  [arXiv:1302.4733 [hep-th]].

\bibitem{RST}
C.~G.~Callan, Jr., S.~B.~Giddings, J.~A.~Harvey and A.~Strominger, Phys.\ Rev.\ D {\bf 45}, 1005 (1992)  [hep-th/9111056]; 
J.~G.~Russo, L.~Susskind and L.~Thorlacius, Phys.\ Rev.\ D {\bf 46}, 3444 (1992) [hep-th/9206070]; 
J.~G.~Russo, L.~Susskind and L.~Thorlacius, Phys.\ Rev.\ D {\bf 47}, 533 (1993)  [hep-th/9209012]. 

\bibitem{black star} 
C.~Barcelo, S.~Liberati, S.~Sonego and M.~Visser, Phys.\ Rev.\ D {\bf 77}, 044032 (2008) [arXiv:0712.1130 [gr-qc]].

\bibitem{firewall} 
  S.~L.~Braunstein, S.~Pirandola and K.~Zyczkowski, Phys.\ Rev.\ Lett.\  {\bf 110}, no. 10, 101301 (2013) [arXiv:0907.1190 [quant-ph]]; 
  A.~Almheiri, D.~Marolf, J.~Polchinski and J.~Sully, JHEP {\bf 1302}, 062 (2013) [arXiv:1207.3123 [hep-th]]; 
  L.~Susskind, arXiv:1301.4505 [hep-th]; 
  J.~Maldacena and L.~Susskind, Fortsch.\ Phys.\  {\bf 61}, 781 (2013)  [arXiv:1306.0533 [hep-th]].

\bibitem{Duff1} 
  M.~J.~Duff,
  Nucl.\ Phys.\ B {\bf 125}, 334 (1977).
  

\bibitem{anomaly}
S.~M.~Christensen and S.~A.~Fulling, Phys.\ Rev.\ D {\bf 15}, 2088 (1977); 
E.~Mottola and R.~Vaulin, Phys.\ Rev.\ D {\bf 74}, 064004 (2006) [gr-qc/0604051].

\bibitem{anomaly2}
S.~P.~Robinson and F.~Wilczek, Phys.\ Rev.\ Lett.\  {\bf 95}, 011303 (2005) [gr-qc/0502074]; 
S.~Iso, H.~Umetsu and F.~Wilczek, Phys.\ Rev.\ Lett.\  {\bf 96}, 151302 (2006) [hep-th/0602146]; 
S.~Iso, H.~Umetsu and F.~Wilczek, Phys.\ Rev.\ D {\bf 74}, 044017 (2006) [hep-th/0606018].

\bibitem{spectral}
  D.~M.~Capper and M.~J.~Duff, Nucl.\ Phys.\ B {\bf 82}, 147 (1974); 
  A.~Cappelli, D.~Friedan and J.~I.~Latorre, Nucl.\ Phys.\ B {\bf 352}, 616 (1991).

\bibitem{BD}
N.~D.~Birrell and P.~C.~W.~Davies, \textit{Quantum Fields in curved space} (Cambridge Univ. Press, Cambridge, 1982). 

\bibitem{Hawking2}
  S.~W.~Hawking,
  Phys.\ Rev.\ D {\bf 14} (1976) 2460.  

\bibitem{Bekenstein} 
  J.~D.~Bekenstein,
  Phys.\ Rev.\ D {\bf 7}, 2333 (1973).  

\bibitem{works}
 V.~P.~Frolov and G.~A.~Vilkovisky, Phys.\ Lett.\ B {\bf 106}, 307 (1981); 
 C.~R.~Stephens, G.~'t Hooft and B.~F.~Whiting, Class.\ Quant.\ Grav.\  {\bf 11}, 621 (1994)  [gr-qc/9310006]; 
  A.~Ashtekar and M.~Bojowald, Class.\ Quant.\ Grav.\  {\bf 22}, 3349 (2005) [gr-qc/0504029]; 
   S.~A.~Hayward, Phys.\ Rev.\ Lett.\  {\bf 96}, 031103 (2006) [gr-qc/0506126]; 
 A.~Ashtekar, V.~Taveras and M.~Varadarajan, Phys.\ Rev.\ Lett.\  {\bf 100}, 211302 (2008) [arXiv:0801.1811 [gr-qc]]; 
 S.~Hossenfelder, L.~Modesto and I.~Premont-Schwarz, Phys.\ Rev.\ D {\bf 81}, 044036 (2010) [arXiv:0912.1823 [gr-qc]]; 
  C.~Rovelli and F.~Vidotto, arXiv:1401.6562 [gr-qc]; 
 V.~P.~Frolov, JHEP {\bf 1405}, 049 (2014) [arXiv:1402.5446 [hep-th]]; 
   Y.~Liu, D.~Malafarina, L.~Modesto and C.~Bambi, Phys.\ Rev.\ D {\bf 90}, 044040 (2014) [arXiv:1405.7249 [gr-qc]]; 
 J.~M.~Bardeen, arXiv:1406.4098 [gr-qc]; 
  H.~M.~Haggard and C.~Rovelli, arXiv:1407.0989 [gr-qc]; 
  C.~Barcelo, R.~Carballo-Rubio and L.~J.~Garay, arXiv:1407.1391 [gr-qc]; 
  C.~Barcelo, R.~Carballo-Rubio, L.~J.~Garay and G.~Jannes, arXiv:1409.1501 [gr-qc].

\bibitem{H-E}
 S.~W.~Hawking and G.~F.~R.~Ellis, \textit{The Large Scale Structure of Space-Time} (Cambridge Univ. Press, Cambridge, 1975). 

\end{thebibliography}
\end{document}